\documentclass[pre,aps,twocolumn,superscriptaddress,floatfix]{revtex4-1}
\usepackage{float}
\usepackage{amsmath, amssymb, amsthm, amsfonts}
\usepackage[applemac]{inputenc}
\usepackage{color}

\usepackage{graphicx}

\begin{document}

\title{Predicting brain evoked response to external stimuli from temporal correlations of spontaneous activity}

\author{A. Sarracino}
\email{alessandro.sarracino@unicampania.it}
\affiliation{Engineering Department, University of Campania ``Luigi Vanvitelli'', 81031 Aversa (Caserta), Italy}

\author{O. Arviv}
\email{oshrit.arviv@biu.ac.il}
\affiliation{Department of Cognitive and Brain Sciences, Ben-Gurion University of the Negev, Beer-Sheva, Israel}
\affiliation{The Inter-Faculty School for Brain Sciences, Zlotowski Center for Neuroscience, Ben-Gurion University of the Negev, Beer-Sheva, Israel}

\author{O. Shriki}
\email{shrikio@bgu.ac.il}
\affiliation{Department of Cognitive and Brain Sciences, Ben-Gurion University of the Negev, Beer-Sheva, Israel}
\affiliation{The Inter-Faculty School for Brain Sciences, Zlotowski Center for Neuroscience, Ben-Gurion University of the Negev, Beer-Sheva, Israel}
\affiliation{Department of Computer Science, Ben-Gurion University of the Negev, Beer-Sheva, Israel}

\author{L. de Arcangelis}
\email{lucilla.dearcangelis@unicampania.it}
\affiliation{Engineering Department, University of Campania ``Luigi Vanvitelli'', 81031 Aversa (Caserta), Italy}

\begin{abstract}
The relation between spontaneous and stimulated global brain activity
is a fundamental problem in the understanding of brain functions.
This question is investigated both theoretically and experimentally
within the context of
nonequilibrium fluctuation-dissipation relations.
We consider the stochastic coarse-grained Wilson-Cowan
model in the linear noise approximation and compare analytical results
to experimental data from magnetoencephalography (MEG) of human brain.
The short time behavior of the autocorrelation function for
spontaneous activity is characterized by a double-exponential decay,
with two characteristic times, differing by two orders of
magnitude. Conversely, the response function exhibits a single
exponential decay in agreement with experimental data for evoked
activity under visual stimulation.  Results suggest that the brain
response to weak external stimuli can be predicted from the
observation of spontaneous activity and pave the way to controlled
experiments on the brain response under different external
perturbations.
\end{abstract}

\maketitle

\section{Introduction} 

The brain represents one of the most
fascinating systems where several mechanisms at different scales are
deeply intertwined, resulting in a complex behavior.   
One of the main open issues in the understanding of brain functioning is
the relation between spontaneous and evoked activity, namely the
response of the system to external stimuli. In particular, it has been found that the large
variability in the response to repeated presentations of the same
stimulus can be attributed to the ongoing spontaneous
activity~\cite{arieli, fox2, fox}. However, a theoretical framework to formalize this 
question is still lacking and the quantitative connection between the spontaneous
and evoked activity remains unclear~\cite{he, huang}.

Experimental results for temporal correlations of spontaneous brain
activity have been reported in a number of studies. For instance, the
seminal article~\cite{linkenkaer} focused on correlations for
spontaneous alpha oscillations in the healthy human brain, and found a
decay characterized by power-law behavior at long times. More recent
analyses~\cite{murray} measured autocorrelations of spiking activity
fluctuations in several cortical areas of the macaque monkey. For each area an intrinsic timescale is defined from the exponential fit of the autocorrelation, which decays to a nonzero offset value taking into account longer timescales. A hierarchical ordering of intrinsic characteristic times
across areas was evidenced in \cite{honey}.  In particular, a
hierarchy of time scales coupling the dynamics of sensory regions at
different ordering levels, was found in the auditory functional area
\cite{steph}. The observed relation between time-scales suggests a temporal organization across the cerebral cortex.
Other studies on the cortex dynamics, focusing on the
effects of sleep deprivation on time correlations, observed
exponential decays: Sustained wakefulness appears to decrease
correlation timescales in humans~\cite{meisel2}, and cortical
timescales depend on time awake and sleep stage in rats~\cite{meisel}.
Even if spontaneous activity has been found to be highly structured
and to participate in high cognitive functions, its functional role
remains poorly understood. In particular, it is well accepted that
variability is reduced during tasks at different scales, yet there is no
general consensus whether task activation would inflate or reduce
functional connectivity and correlations \cite{fere, cole}. The
question we address here is to provide a theoretical formulation for
the relation between the response and the correlation function of rest
activity by means of a simple neuronal model suitable for an
analytical approach.

The relation between ongoing and stimulated activity can be addressed 
theoretically within the general framework of statistical physics, 
by means of the
fluctuation-dissipation relations, connecting the
spontaneous fluctuations of a system with the response function to
external perturbations. Recently, these relations have been extended
beyond the context of standard thermodynamics, to nonequilibrium
systems, such as active and biological matter~\cite{vulp, cugli, seif, baiesi, sarra, sarra2}.
These relations hold for a wide variety of physical systems, not necessarily at the critical point. Their deep predictive power relies on the possibility
to estimate the response of a system from
the observation of its spontaneous fluctuations. However, in order to
write explicit relations, a theoretical model describing the system dynamics at the 
scale of interest is needed. To this extent, a useful approach to describe
the activity of a large number of neurons is provided by the coarse-grained
Wilson-Cowan model (WCM)~\cite{wc0, wc1}.
The stochastic version of this model can describe the fluctuations of 
active excitatory and inhibitory neuron populations via two Langevin equations, whose
dynamics is coupled through a feed-forward term, and can reproduce the 
statistics of burst activity \cite{wc}. 
Recent studies focused on dynamical stability properties of this model
revealing a rich dynamical behavior, see
for instance~\cite{nega, livi, burioni}.

In this work, we address the fundamental problem of unveiling the
quantitative relation between spontaneous and stimulated brain
activity. We compare the analytical expression for correlations and response
function in the WCM with spontaneous and stimulated
brain activity measured via magnetoencephalography (MEG).  We
measure the time correlation functions of spontaneous activity
in several healthy subjects, finding a temporal behavior mainly characterized
by a double-exponential decay.  This two-timescale decay
observed in experiments is in good agreement with the analytical prediction
of the WCM. We also calculate, in the linear response regime, 
the response function to an external stimulation and 
compare the result to MEG data of evoked activity, obtained by applying a visual
stimulation (pictures of faces) to the same subjects. Our study
enlightens how some properties of the induced brain dynamics can be
predicted from the observation of spontaneous activity in the absence
of stimuli.

The paper is organized as follows. In Sec. \ref{wcmodel} we briefly recall the definition of the WCM,
focusing on its linearized version, which is expected to hold in the large size limit.
Then, in Sec. \ref{megdata}, we report the experimental results for the spontaneous activity autocorrelation functions from MEG data.
In Sec. \ref{respsec} we derive the fluctuation-dissipation relation for the WCM
and compare its predictions with experimental data for evoked activity. Finally some conclusions are drawn in
the last section. Appendices A, B, and C report details on the experimental procedure and on analytical calculations.

\section{Linearized Wilson-Cowan model} 
\label{wcmodel}

In order to provide a
theoretical framework to understand the relation between spontaneous
and evoked brain activity, we consider the linearized version of the
stochastic coarse-grained WCM, which allows us to describe the systems dynamics at meso- and macroscales \cite{wc1}.
This model describes the excitatory and
inhibitory neuron population dynamics, which evolve according to a
Master Equation for the probability $p_{k,l}(t)$ to have $k$
excitatory and $l$ inhibitory neurons \emph{active} at time $t$~\cite{wc}. The
total number of neurons is $N_E+N_I$, where $N_E$ and $N_I$ are the
populations of excitatory and inhibitory neurons, respectively. 
Assuming $N_E=N_I=N$, for large $N$ the number of active neurons is written as
the sum of a deterministic component $(E,I)=(k/N,l/N)$,
and a stochastic fluctuation $(\xi_E,\xi_I)$, scaled by
$\sqrt{N}$, i.e. $k=NE+\sqrt{N}\xi_E$ and $l=NI+\sqrt{N}\xi_I$.
The Master Equation can be rewritten in terms of these new variables and
expanded in power of $N$ obtaining a Fokker-Planck equation. This leads to the dynamic equations for the deterministic and stochastic components.
Next, introducing the global variables \cite{wc}
\begin{equation}
  \Sigma=(E+I)/2, \qquad
  \Delta=(E-I)/2, 
\end{equation}
in the large $N$ limit and close to the fixed point ($\Sigma_0,\Delta_0=0$) the deterministic components satisfy the equations
\begin{equation}
  \frac{d\Sigma}{dt}=-\alpha \Sigma + (1-\Sigma)f(s)/\tau_0, \qquad
  \frac{d\Delta}{dt}=-\Delta(\alpha+f(s)/\tau_0),
\end{equation}
where $\alpha$ is the decay rate of the active state, $f(s)=\tanh(s)$ for $s>0$
and zero otherwise, $\tau_0=1$ ms is a fixed microscopic time-scale and $s=w_0\Sigma+(w_E+w_I)\Delta + h$. Here 
$w_E$ ($w_I$) is the synaptic strength of the excitatory (inhibitory) population,
$w_0=w_E-w_I$ and $h$ a small external input. We notice that $\Sigma$ represents the global neuronal activity, whereas $\Delta$ measures the unbalance in activity between the excitatory and the inhibitory population.
The deterministic equations have the unique stable solution
$(\Sigma_0,0)$. Note that the fixed point $\Delta_0=0$ represents the
condition of balance between excitation and inhibition, as found for
spontaneous activity of neuronal systems in healthy state \cite{lomb}.

Concerning the fluctuations of $\Sigma$ and $\Delta$, $\xi_\Sigma$ and $\xi_\Delta$, in the large $N$ limit (the linear noise approximation) and close to the fixed point ($\Sigma_0$, $\Delta_0=0$), they satisfy~\cite{wc}
\begin{equation}\label{model}
  \frac{d}{dt}
\left( \begin{array}{c}
\xi_\Sigma \\
\xi_\Delta \end{array} \right) = \left( \begin{array}{cc}
-1/\tau_1 & w_{ff}\\
0 & -1/\tau_2 \end{array} \right)\left( \begin{array}{c}
\xi_\Sigma \\
\xi_\Delta \end{array} \right) + \sqrt{\alpha \Sigma_0}\left( \begin{array}{c}
\eta_\Sigma \\
\eta_\Delta \end{array} \right),
\end{equation}
where $\eta_\Sigma$ and $\eta_\Delta$ are zero average,
delta-correlated white noises with unitary variance, and
$1/\tau_1=\alpha+f(s_0)/\tau_0-(1-\Sigma_0)w_0f'(s_0)/\tau_0$,
$w_{ff}=(1-\Sigma_0)(w_E+w_I)f'(s_0)/\tau_0$, $1/\tau_2=\alpha+f(s_0)/\tau_0$, with
$s_0=w_0\Sigma_0+h$.  The off-diagonal term $w_{ff}$ is called hidden
feed-forward term and plays a central role because it rules the
coupling between the global activity and the unbalance between the activities of excitatory and inhibitory populations~\cite{mm}.
In particular, a fluctuation in $\Delta$ affects the temporal evolution of $\Sigma$
but not vice-versa.

Eqs.~(\ref{model}) are two coupled linear Langevin equations, the
dynamics of which can be easily solved analytically~\cite{risken}.  In
particular, the solution in vectorial form reads
\begin{equation}\label{sol}
\boldsymbol{\xi}(t)=e^{{\cal M} t}\boldsymbol{\xi}(0)+\sqrt{\alpha \Sigma_0}\int_0^t e^{{\cal M}(t-t')}\boldsymbol{\eta}(t')dt',
\end{equation}
where ${\cal M}$ is the coupling matrix appearing in Eq.~(\ref{model}).
Therefore, we evaluate the correlation matrix at stationarity
\begin{equation}\label{corr0}
{\cal C}_{ij}(t)\equiv \langle \xi_i(t)\xi_j(0)\rangle=(e^{{\cal M}t}\sigma)_{ij},
\end{equation}
where $\langle\cdots \rangle$ denote average over noise, $i,j=(\Sigma,\Delta)$ 
and $\sigma$ is the
covariance matrix which is a known function of the model
parameters (see Appendix C).
The specific case of
the autocorrelation for the total activity $\Sigma$,
${\cal C}_{\Sigma\Sigma}\equiv\langle \xi_\Sigma(t)\xi_\Sigma(0)\rangle$,
takes the form
\begin{eqnarray}  \label{corr}
  {\cal C}_{\Sigma\Sigma}(t)&=&\frac{\alpha \Sigma_0}{2(\tau_1^{-2}-\tau_2^{-2})} \\
  &\times&\left(\frac{\tau_1^{-2}-\tau_2^{-2}-w_{ff}^2}{\tau_1^{-1}}e^{-t/\tau_1}+\frac{w_{ff}^2}{\tau_2^{-1}}e^{-t/\tau_2}\right).\nonumber
\end{eqnarray}
Therefore, the behavior of the correlations of the active neuron
population is characterized by a double-exponential decay, with two
characteristic times $\tau_1$ and $\tau_2$, which are a function of the model parameters 
as specified above. Similar double exponential decay is also found for the 
${\cal C}_{\Sigma\Delta}(t)$, whereas a single exponential function with characteristic 
time $\tau_2$ is obtained for ${\cal C}_{\Delta\Sigma}(t)$ and  ${\cal C}_{\Delta\Delta}(t)$ (see Appendix C).

\begin{figure}[!tb]
	\centering
	\includegraphics[width=0.8\columnwidth,clip=true]{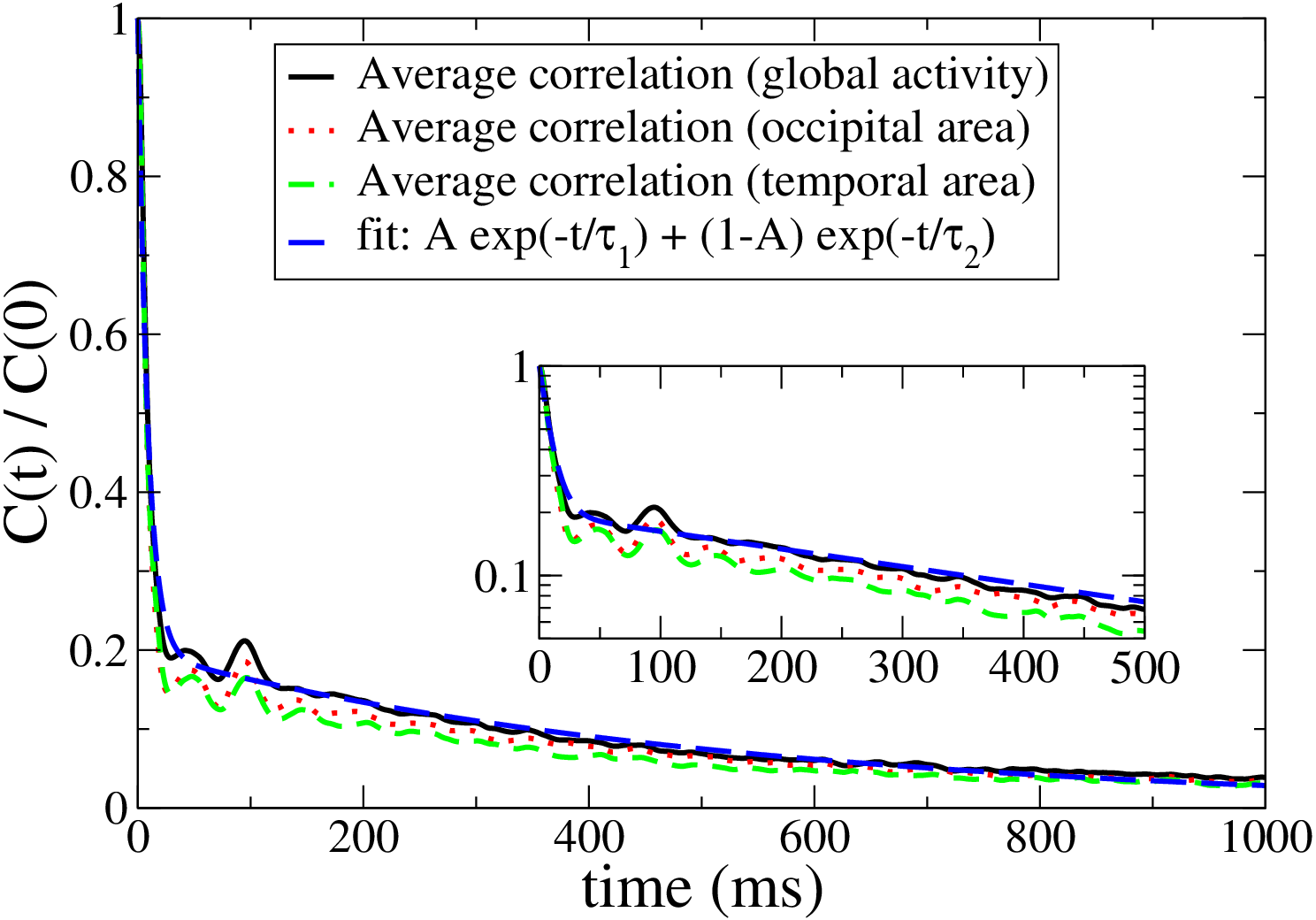}
	\caption{Normalized time autocorrelation of the spontaneous
          activity.  in linear (main) and semi-log (inset) scale for
          global activity and for the two functional areas. Lines
          represent the average over 14 samples (2 per subject). The
          blue thick dashed line represents the best fit curve
          $f(t)=A\exp(-t/\tau_1)+(1-A)\exp(-t/\tau_2)$, with
          parameters $A=0.8\pm 0.1$, $\tau_1=8.8 \pm 1.5$ ms and
          $\tau_2=515\pm 200$ ms.}
	\label{fig1}
\end{figure}

\section{Time correlation of spontaneous activity from MEG data} 
\label{megdata}

The
variable $\Sigma$ describes the brain spontaneous activity and
therefore allows for a comparison with experimental data obtained from
MEG measurements. We analyse the global signal, by summing the signals at all sensors, the signal from sensors monitoring the visual area in the occipital lobe and the signal from sensors monitoring the temporal lobes, which include the area specialized in face recognition, e.g., the Fusiform Face Area (FFA).
Spontaneous activity was recorded from 7 healthy human subjects with a
whole-head 248-channel magnetometer array (4-D Neuroimaging, Magnes
3600WH) in the MEG facility at the EMBI Unit, Bar-Ilan University,
Israel.  The MEG signal was recorded at a sampling rate of 1017.25 Hz
and offline band-pass filtered between 0.8 and 80 Hz as well as
underwent a cleaning procedure of potential artefacts. The absolute
amplitudes were summed across the recorded channels to give a global
rest activity signal.  More details can be found in Appendix A and
in~\cite{arviv, arviv2}.  We compute the normalized autocorrelation
function
\begin{equation}
\frac{C(t)}{C(0)}=\frac{\sum_{s=1}^{N-t}(|x(s)|-\mu)(|x(s+t)|-\mu)}{\sigma_0}
\end{equation}
from time series (of total length about 4 minutes for each subject) of
the absolute value $\{|x(t)|\}$ of rest activity signal, with mean
$\mu$ and standard deviation $\sigma_0$. We focus here on the short
time behavior (up to one second), where the experimental data show a
decay mainly characterized by two typical times (see
Fig.~\ref{fig1}). In this study we neglect oscillations at very short
time scales, which probably correspond to $\alpha$ brain rhythm, and
consider the global decay behavior.  Analyzing 7 subjects (2 data sets
for each subject), we observe that this complex behavior is quite
stable (see Table in Appendix B) and can be fitted
by Eq.~(\ref{corr}). Moreover, the same functional behavior is found
for the global signal and for the signal averaged only on 
sensors from temporal and occipital lobes, corresponding to areas involved in processing of the stimuli. The double exponential
behavior is also detected for data at single sensors in these two
areas, where the autocorrelation function, as expected, exhibits
larger statistical fluctuations (see Appendix B).  The average values
of the characteristic times are found to be $\tau_1= 8.8 \pm 1.5$ ms
and $\tau_2= 515 \pm 200$ ms, differing by almost two orders of
magnitude. The short time scale, causing an abrupt decay in the
correlation function, is of the order of magnitude of synaptic time
scales as well as single neuron time scales, such as the refractory
period \cite{som}. Conversely, the long characteristic time is
compatible with the time scale of low-frequency rhythms, as the
$\theta$ rhythm.  Interestingly, the long characteristic time $\tau_2$
controls the single exponential decay of the correlations
${\cal C}_{\Delta\Sigma}(t)$ and ${\cal C}_{\Delta\Delta}(t)$ (see Appendix C). This
suggests that a fluctuation in $\Sigma$ or $\Delta$ affects the
temporal behavior of $\Delta$ itself over a long timescale, implying a
slow recovery of the balance condition if the system is moved away
from the fixed point $\Delta_0=0$.  From the fitting procedure we also
obtain an estimation for the feed-forward coefficient $w_{ff}=0.008\pm
0.0035$ ms$^{-1}$, which characterizes the coupling between excitatory
and inhibitory neuron populations.
Finally, we note that, at longer times
($t\gtrsim 2$ s), slower decay (power-law) of the autocorrelation function can take place. This functional decay cannot be obtained by a linearized model. 

\section{Predicting response from spontaneous activity via the Fluctuation-Dissipation Theorem} 
\label{respsec}

The present theoretical framework
allows us to address the fundamental question of the relation between 
spontaneous and evoked activity. We can evaluate
the linear response function to a weak external perturbation, defined as
\begin{equation}
{\cal R}_{ij}(t)\equiv\frac{\overline{\delta \xi_i(t)}}{\delta \xi_j(0)},
  \end{equation}
where $i,j=(\Sigma,\Delta)$. This represents the average response of
the variable $ \xi_i(t)$ at time $t$ to an impulsive perturbation
applied to the variable $\xi_j$ at time 0, and $\overline{\cdot}$
denotes non-stationary averages over many realizations.  
From Eq.(\ref{sol}) one immediately obtains the response matrix
\begin{equation}\label{resp0}
{\cal R}(t)=e^{{\cal M}t} \qquad \textrm{for}\qquad t>0.
  \end{equation}
From Eq.~(\ref{corr0}), we obtain a fluctuation-dissipation
relation
\begin{equation}
{\cal R}(t)={\cal C}(t)\sigma^{-1},
  \end{equation}   
connecting the linear response with spontaneous
fluctuations~\cite{vulp}. In particular, since we are interested in the
response of the total activity $\Sigma$ to a small perturbation of $\Sigma$
itself, the response function reads
\begin{equation}
  {\cal R}_{\Sigma\Sigma}(t)=(\sigma^{-1})_{11}\langle \xi_\Sigma(t)\xi_\Sigma(0)\rangle + (\sigma^{-1})_{12}\langle \xi_\Sigma(t)\xi_\Delta(0)\rangle,
  \label{fdt}
\end{equation}
where the explicit form of the inverse matrix $\sigma^{-1}$ is reported in Appendix C.
The relevance of relation~(\ref{fdt}) relies on the fact that it
allows us to get information on the response of the
system to an external perturbation by simply looking at its
unperturbed dynamics.  In general, as evident from Eq.~(\ref{fdt}),
both the autocorrelation and the cross-correlation are required to
reconstruct the response to a weak external stimulus. This is due to
the presence of nonequilibrium conditions breaking detailed balance,
as noted in several systems (see for instance~\cite{vulp,svgp,cpv}).

Let us note however, that, in our particular case, due to the upper
triangular form of the matrix ${\cal M}$ in Eq.~(\ref{model}), the response
function of the model takes the simple exponential form
\begin{equation}
  {\cal R}_{\Sigma\Sigma}(t)=\exp(-t/\tau_1),
  \label{resp}
\end{equation}
where the characteristic time $\tau_1$ is the same ruling the
short-time behavior of the correlation function. This is consistent
with Eq.~(\ref{fdt}) because one can easily check that the
cross-correlation appearing in Eq.~(\ref{fdt}) exactly cancels the
term $\propto \exp(-t/\tau_2)$ in Eq.~(\ref{corr}), giving the
single-exponential decay for the response function (see Appendix C). 
Therefore, in this
approach, measurements of the spontaneous fluctuations in the global brain
activity $\Sigma$ alone could provide a prediction for the system response.
Concerning the other response functions, we notice that 
${\cal R}_{\Sigma\Delta}(t)$ shows a double exponential behavior, whereas,
as expected from the triangular form of ${\cal M}$, ${\cal R}_{\Delta\Sigma}(t)=0$ 
and ${\cal R}_{\Delta\Delta}(t)=\exp(-t/\tau_2)$ (see Appendix C).

\begin{figure}[!tb]
\centering
\includegraphics[width=0.8\columnwidth,clip=true]{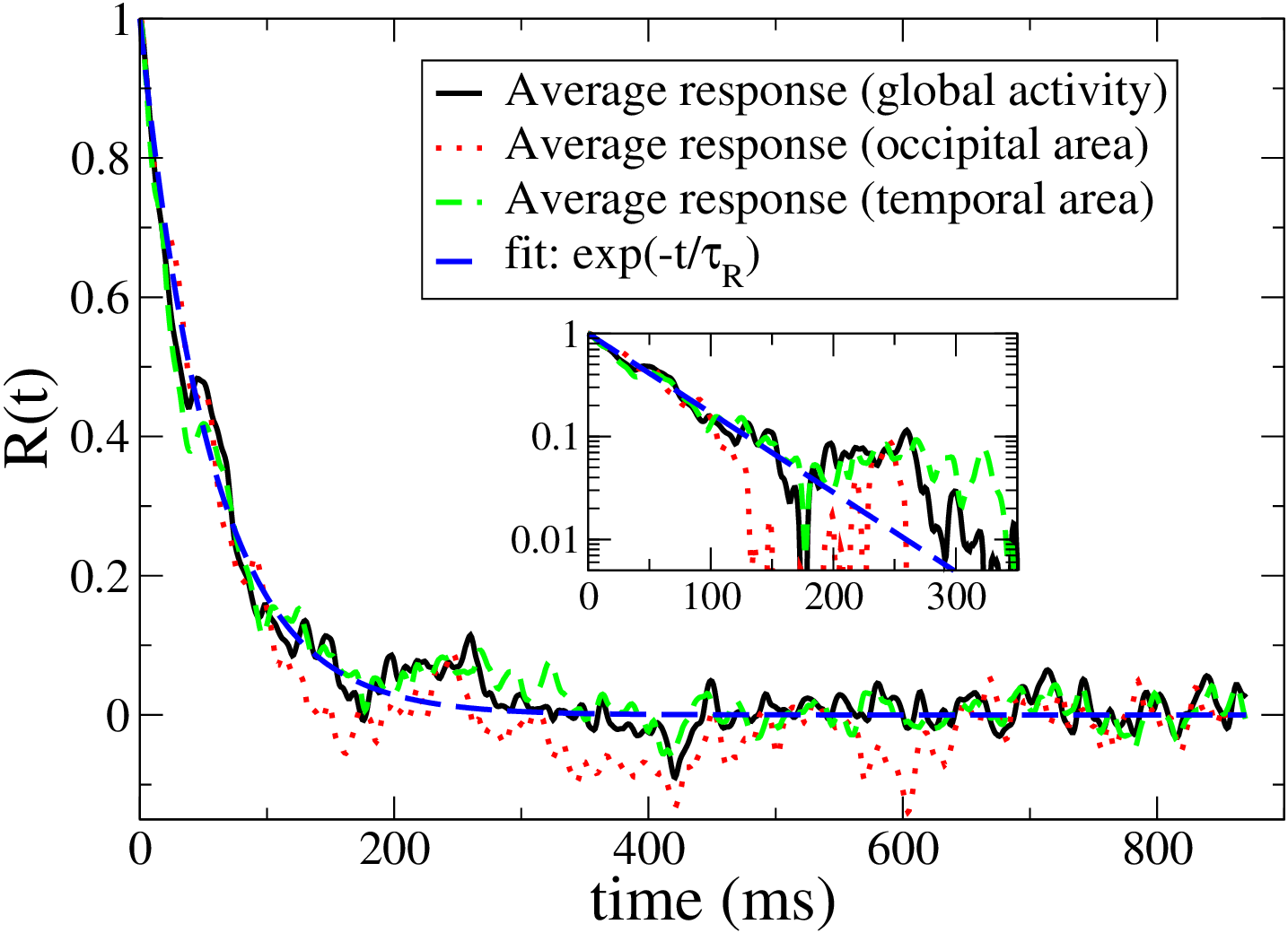}
\caption{(Rescaled and shifted) average response function for the global signal, the signal
from the occipital and the temporal lobes.  The typical time for the response
  function from an exponential fit decay, according to
  Eq.~(\ref{resp}), is $\tau_R=55 \pm 25$ ms. In the inset the same data are plotted in semi-log scale.}
\label{fig2}
\end{figure}

The response function derived by the analytical calculation characterizes the behavior close-in-time to the stimulus application. In order to fairly compare the analytical prediction to experimental data, the experimental perturbation has
to be sufficiently weak to remain in the linear regime.
Face processing is one of the most studied cognitive abilities. Humans are considered experts in face processing,  which accordingly does not involve long-lasting cognitive engagement, potentially mimicking a small perturbation to the system. The evoked response is fast and of relatively short (typically characterized by 100, 170 and 250 ms evoked response fields (ERFs), and some late components at 400 and 600 ms) and the return to baseline in both power and spectral content occurs prior to the 1 sec limit \cite{arviv, arviv2}. We measured the brain activity of the same 7 participants evoked by visual stimuli, by showing a series of pictures of human faces. Each stimulus is presented for 1 sec, with varying inter stimulus intervals (larger than 1 sec). Further details are reported in Appendix A.

In Fig.~\ref{fig2} we show the response function averaged over the 7
subjects for the global signal and for the signal averaged only on 
sensors from temporal and occipital lobes, corresponding to areas involved in processing of the stimuli (see Appendix B for the response function for each subject and at single
sensor).  Since each subject shows a different latency activity
associated to each stimulus which cannot be accounted for by
analytical predictions, experimental data for the response have been
shifted in time and rescaled in amplitude in order to have that the
maximum in each dataset is reached at time zero and its value is
normalized to 1. We have to stress that the applied perturbation
cannot be considered impulsive, as in the analytical definition, and
moreover the system takes a certain time to develop the response
signal, as shown in Appendix B. However, experimental data on
evoked activity qualitatively confirm the single exponential decay
predicted by the WCM for all datasets. Indeed, imposing a fit with a
double exponential leads in all cases to an amplitude close to zero
for the second exponential decay. The exponential behavior is also
detected for data at single sensors, where the response function, as
expected, exhibits larger statistical fluctuations (see Appendix B).
The averaged fitted value for the characteristic time $\tau_R=55\pm
25$ms, which is of order of magnitude comparable to $\tau_1$ (see Table
in Appendix B for the characteristic time values in each dataset). The
quantitative difference between these two timescales could be due to a
number of factors, as the properties of the stimulation, which in the
theoretical approach is supposed to be instantaneous and small. The
real stimulation requires a time delay before the decay sets in,
raising the need to identify the initial time to fit the response
function.
Fluctuations in experimental conditions then provide a wider range of
characteristic times that, averaged over trials, lead to a larger
$\tau_R$. Random fluctuations also hide a possible correlation between
$\tau_1$ and $\tau_R$ across subjects.  Response experiments under
different stimulation protocols and controlled conditions are required
to test more quantitatively the timescale agreement.  The robust functional behavior found for global signals, functional areas signals and data at single sensor,  is an indication of scale-invariance, namely the correlation and the response function do not depend on the sample size, therefore evidencing the self-similar properties of the system. This
observation could be the outcome of the coarse-grained measure of the
activity by MEG, since sensors are spaced at $\sim 2$cm and
therefore the signal at a single sensor is already representative of
the activity of a large population of neurons. We are aware that at a
finer scale the response to a stimulus has a complex spatio-temporal
organization, however this information cannot be accounted for by our
analytical approach.

Let us finally observe that within this approach one can derive another
relevant quantity characterizing the nonequilibrium behavior of the
neuronal system, namely the entropy production rate, which characterizes how much the system is far from equilibrium~\cite{cpv}
\begin{equation}
\frac{1}{t}\langle W_t\rangle=\frac{2 w_{ff}}{\alpha \Sigma_0(\tau_1^{-1}+\tau_2^{-1})}.
\end{equation}
This makes clear that the nonequilibrium source in this model is the
presence of the feed-forward term $w_{ff}$, which couples the fluctuations in the two variables $\Sigma$ and $\Delta$.

\section{Conclusions} 

We have proposed a theoretical approach to predict the relation between
global spontaneous and evoked brain activity, based on the linear noise approximation of the WCM, which allows for analytical calculations. The theoretical approach relies on two main assumptions: The model is two-dimensional and parameters are tuned to achieve balanced excitation and inhibition, which leads to the fixed point $\Delta=0$.
The comparison with MEG data confirms that temporal correlations of spontaneous fluctuations are mainly characterized by a
double-exponential decay, with two well separated typical times
$\tau_1\simeq 10$ ms and $\tau_2 \simeq 500$ ms. The short time scale
is related to synaptic timescales, while the long
one is compatible with slow brain rhythms. 
This analytical approach is therefore able to rationalize experimental results for the correlation and response functions providing a coherent framework for the dual functional behavior found experimentally. An important aspect of our results is that the functional behavior found in MEG data is scale-invariant, with stable value of the characteristic times
across scales, from the cm scale (single sensor data) to the entire brain (global signal). 
This observation suggests that at first approximation the linear model well accounts for
the brain behavior at different scales, even if the connectome has a complex modular structure.

The presence of a single exponential decay with the short
characteristic time for the response function is fully coherent with
the efficient performance of the human brain in visual tasks. Indeed,
it is well known that the human eye can appreciate about 10 images per
second, resulting in a temporal resolution of about 100 ms \cite{jain}. A larger
relaxation time, of the order of $\tau_2 \simeq 500$ ms, would imply
an overlap in the response to close-in-time stimulations, affecting
the performance. We deem that future experiments on evoked activity,
properly designed to the application of the fluctuation-dissipation
theorem, could shed new light on the brain functionality at large
scale, with strong impact in neurobiology and
neuroscience. Moreover, this result opens the way to numerical studies
implementing microscopic models, as integrate and fire neuronal
models, on complex networks able to investigate in detail the role of
the network structure and of different temporal scales in
neurocognitive behavior.

\begin{acknowledgments}
LdA would like to thank MIUR project PRIN2017WZFTZP for financial support. AS acknowledges support from MIUR project PRIN201798CZLJ. LdA and AS acknowledge support from Program
  (VAnviteLli pEr la RicErca: VALERE) 2019 financed by the University
  of Campania ``L. Vanvitelli''.
\end{acknowledgments}

\appendix

\section{Experimental procedure, Data acquisition and Preprocessing}

Spontaneous resting state activity and stimulus-evoked response were
recorded from healthy human subjects (n = 7, age= $22\pm 3$ years) in
the MEG facility at the EMBI Unit, Bar-Ilan University, Israel. The
participants gave their informed consent and were financially
compensated for their effort. The study was approved by the Bar-Ilan
University ethics committee, in accordance with the relevant
guidelines and regulations.  Neuromagnetic brain activities were
recorded with a whole-head, 248-channel magnetometer array (4-D
Neuroimaging, Magnes 3600WH) in a dimly-lit magnetically-shielded
room, as participants laid supine. In order to rule out head movements
throughout the recordings, head localization measurements were
performed before and after each experiment. Head position and shape
were determined by Pollhemus FASTTRAK digitizer and five coils
attached to the participant's head, measuring position relative to the
MEG sensors.  The MEG was recorded at a sampling rate of 1017.25 Hz
and analog band-pass filtered online at 0.1-400Hz. Reference coils
were used to remove environmental noise. Accelerometers (Bruel and
Kjaer) attached to the gantry were used to remove vibration noise. The
50-Hz signal from the power outlet was recorded by an additional
channel and the average power-line response to a power cycle was
subtracted from every MEG sensor \cite{tal}.

E-prime 2.0 (Psychology Software Tools Inc.) was used for experimental
control. During rest, participants were instructed to fixate their
eyes on a fixation cross at the center of a black screen. During
evoked, the stimuli were grey-scale pictures of human faces. Each
stimulus was presented for 1000 ms with inter-stimulus intervals
varying between 1,300 and 1,700 ms. Stimuli were back projected on a
screen placed in front of the subjects, by a video projector situated
outside the room. To each subject, 540 pictures of faces were
presented (details: photographs of 5 different human male models in 9
head posters and emotional expressions, giving 45 pictures and
photographs of 3 different human female models with 3 emotional
expressions, giving 9 pictures; All pictures were repeated randomly 10
times, once in each experimental block, that is 10 blocks). During the
MEG scan, participants completed an oddball gender-detection task,
pressing a response button only when a female face was presented
(16.67\%). This ensured that all 450 face presentation of male models
were task irrelevant. Only these presentations underwent
analysis. Subsequently, among these 450 picture presentations, between
3 and 30 trials per subject were removed during the cleaning procedure
due to artefacts.

Data processing was performed using
Fieldtrip open-source toolbox for Advanced MEG Analysis
\cite{oost}. MEG recordings were first cleaned for line frequency,
building vibration and heartbeats artefacts with an in-house
open-source software based on external cues
\cite{tal}. Stimulus-evoked data were segmented to include the 1 sec
trials as well as an additional 0.2 pre-trial interval and head and
tail of 0.4 sec that were later cut from analysis. All data were
band-pass filtered offline between 0.8 and 80 Hz (zero-phase two-pass
Butterworth IIR filter of order 4 and 53 dB stopband attenuation,
upper band limit was chosen to minimize the effect of muscle
artifacts). Epochs containing a false-positive response or
contaminated by jump in the MEG sensors or muscle artefacts,
displaying variance higher than 3 SD in power above 60 Hz, were
discarded. One mal-functioning MEG sensor was discarded from all
datasets. Independent component analysis (ICA) was performed on the
remaining data, to ensure the removal of eye-movements, blinks and
leftover heartbeats artefacts. ICA components reflecting such
artefacts, as determined by visual inspection of the 2D scalp maps and
time course of that ICA component, were rejected and remaining
components were used to reconstruct the data. For additional
information, see Ref.\cite{arviv}. The resultant absolute amplitude
were summed across the MEG sensors and formed a global signal of the
associated brain activity.

\section{Autocorrelation and response functions for different functional areas and different subjects}

\begin{figure}[!tb]
\centering
\includegraphics[width=0.9\columnwidth,clip=true]{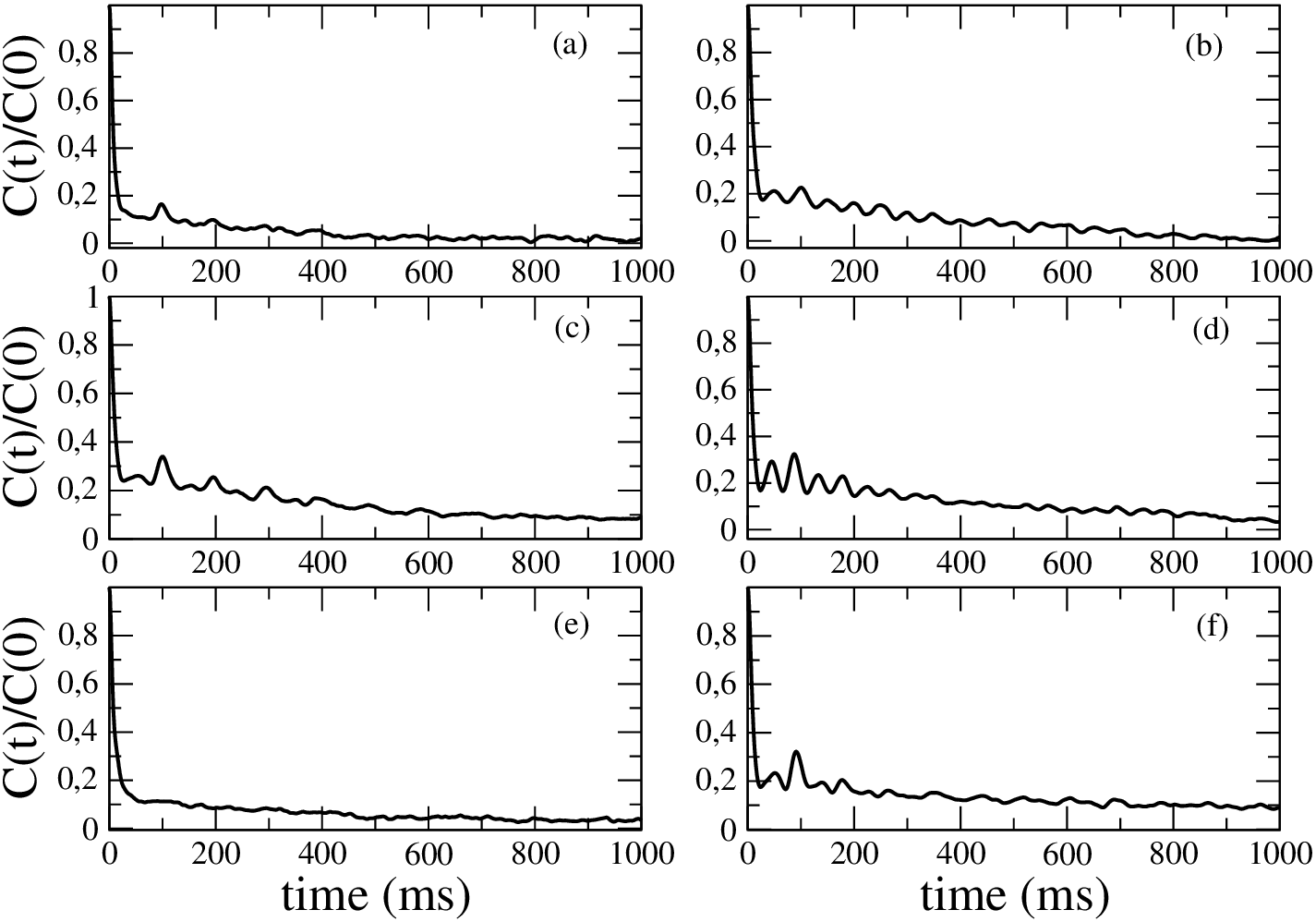}
\caption{Autocorrelation function of the global spontaneous activity
  for subjects 1-6 (a-f).}
\label{corrfig}
\end{figure}

\begin{figure}[!tb]
\centering
\includegraphics[width=0.9\columnwidth,clip=true]{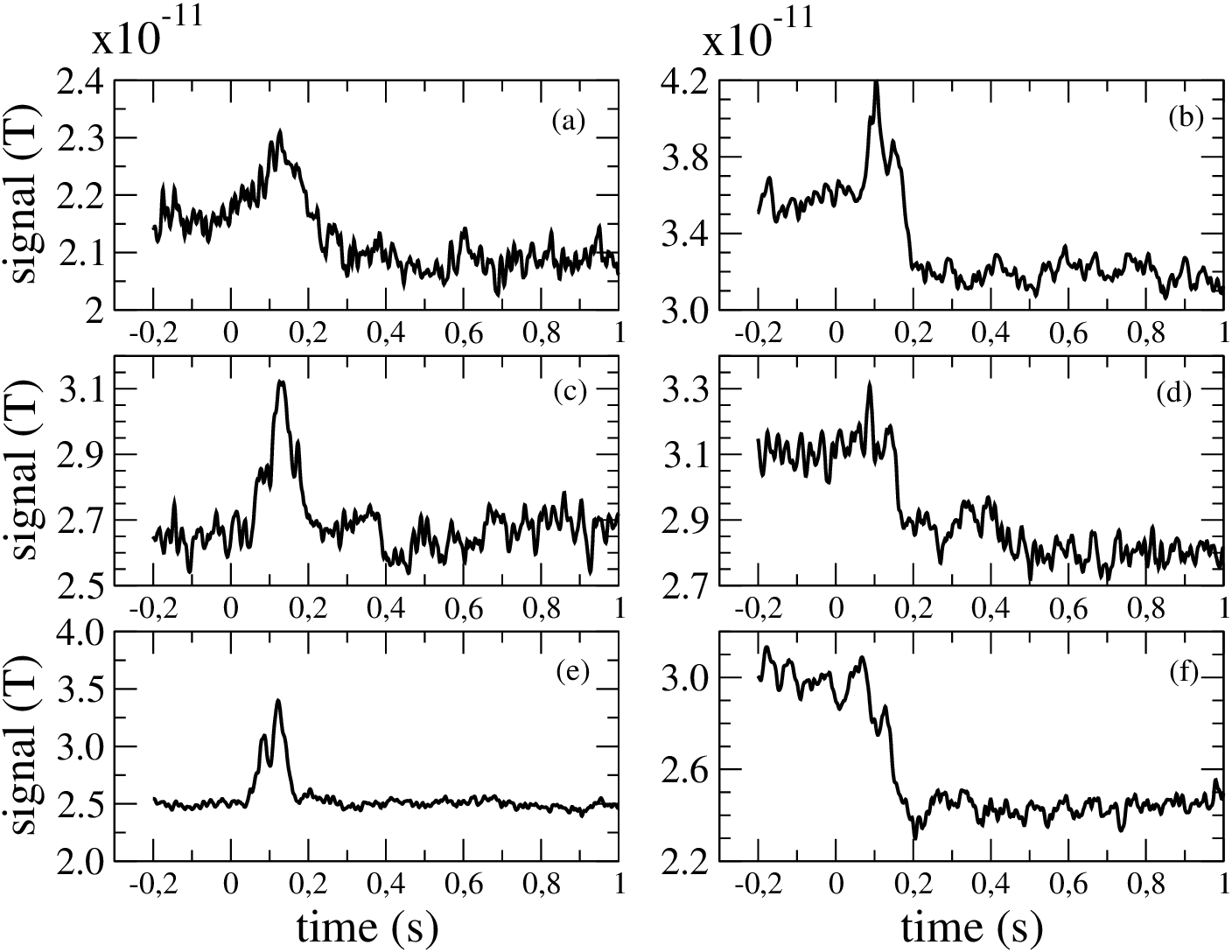}
\caption{Signal of evoked activity for subjects 1-6 (a-f). The visual
  stimulation is applied at time 0.}
\label{respfig}
\end{figure}

In Fig.~\ref{corrfig} and Fig.~\ref{respfig} we report the (global) time
autocorrelations and response functions, respectively, for the several
observed subjects.  

\begin{table*}[t]
\begin{center}
\begin{tabular}{|c||c|c|c||c|c|c||c|c|c||c|c|c||c|c|c|}
 \hline
Subject & $\tau_1$ global & occ. & temp. & $\tau_2$ global & occ. & temp. & $A$ global & occ. & temp. & $\tau_R$ global & occ. & temp.\\
 \hline
 1a  & 7.78  & 7.99 & 8.09 & 259 & 267 & 249 & 0.902 & 0.906 & 0.930 & 83.3 & 45.5 & 96.44 \\        
 1b  & 7.84  & 7.87 & 7.93 & 367 & 394 & 366 & 0.866 & 0.864 & 0.905 & & & \\             
 2a  & 10.82 & 9.45 & 10.1 & 515 & 662 & 598 & 0.789 & 0.756 & 0.84  & 52.6 & 58.7 & 42.2 \\        
 2b  & 8.51  & 8.70 & 8.71 & 400 & 425 & 402 & 0.775 & 0.722 & 0.809 & & & \\
 3a  & 7.75  & 7.60 & 8.22 & 537 & 568 & 740 & 0.757 & 0.849 & 0.890 & 37.0 & 35.7 & 43.5 \\     
 3b  & 8.17  & 7.28 & 8.66 & 698 & 498 & 881 & 0.725 & 0.754 & 0.772 & & & \\
 4a  & 7.65  & 6.42 & 7.20 & 437 & 661 & 591 & 0.832 & 0.874 & 0.921 & 90.9 & 80.4 & 108\\  
 4b  & 7.37  & 6.37 & 6.68 & 582 & 689 & 544 & 0.752 & 0.747 & 0.794 & & & \\
 5a  & 10.65 & 8.80 & 10.6 & 442 & 455 & 439 & 0.742 & 0.897 & 0.878 & 22.7 & 27.2 & 23.5 \\  
 5b  & 9.79  & 7.34 & 8.43 & 637 & 633 & 674 & 0.877 & 0.937 & 0.934 & & & \\   
 6a  & 8.29  & 8.28 & 8.08 & 1049 & 722 & 686 & 0.793 & 0.805 & 0.741 & 52.6  & 37.8 & 54.1 \\   
 6b  & 7.04  & 7.52 & 7.04 & 258  & 409 & 244 & 0.610  & 0.751 & 0.586 & & & \\
 7a  & 10.84 & 9.32 & 9.34 & 691  & 887 & 767 & 0.894 & 0.922 & 0.921 & 41.7 & 128.7 & 30.2 \\ 
 7b  & 10.77 & 8.93 & 9.76 & 815  & 890 & 1314 & 0.855 & 0.872 & 0.888 & & & \\
 \hline
 Average & 8.8$\pm$1.5  & 8.0$\pm$1 & 8.5$\pm$1 & 550$\pm$200 & 583$\pm$190 & 607$\pm$280 & 0.8$\pm$0.1 & 0.8$\pm$0.1 & 0.8$\pm$0.1 & 55$\pm$25 & 60$\pm$35 & 57$\pm$30 \\
 \hline
\end{tabular}
 \caption{Parameters obtained from the double-exponential fit with the
   function $f(t)=Ae^{-t/\tau_1}+(1-A)e^{-t/\tau_2}$ to experimental
   data of rest activity (two data set for each subject) and from a
   single exponential fit with the function $g(t)=e^{-t/\tau_R}$ for
   evoked activity, for all the subjects, for global activity,
   occipital area and temporal area. Times are measured in ms.}
 \label{tab}
\end{center}
\end{table*}

We also show further data analyses considering, rather than the global
signal from all sensors, 1) the signal from sensors corresponding to visual areas (in the occipital lobe) and  2) the signal from sensors corresponding to the temporal lobes, which include the area specialized in face recognition, e.g., the fusiform face area.
 The later are also the regions that demonstrated the strongest activity. 
 For both data sets, we analysed the temporal
autocorrelations for rest activity and the response in the total
signal from each area and for individual sensors in each area.  We
also present data analysis for the different subjects.

In Tab.~\ref{tab} we report the values of the parameters obtained from
the fit of the theoretical predictions of the linearized Wilson-Cowan
to the experimental data, for rest and evoked activity, for global
activity and for occipital and temporal area. For each subject we have
two datasets of spontaneous activity and one data set for evoked
activity. Note that the values of $\tau_1$ are quite stable within the
set of subjects, while the values of $\tau_2$ show a larger
variability.

\begin{figure}[!t]
\centering
\includegraphics[width=0.9\columnwidth,clip=true]{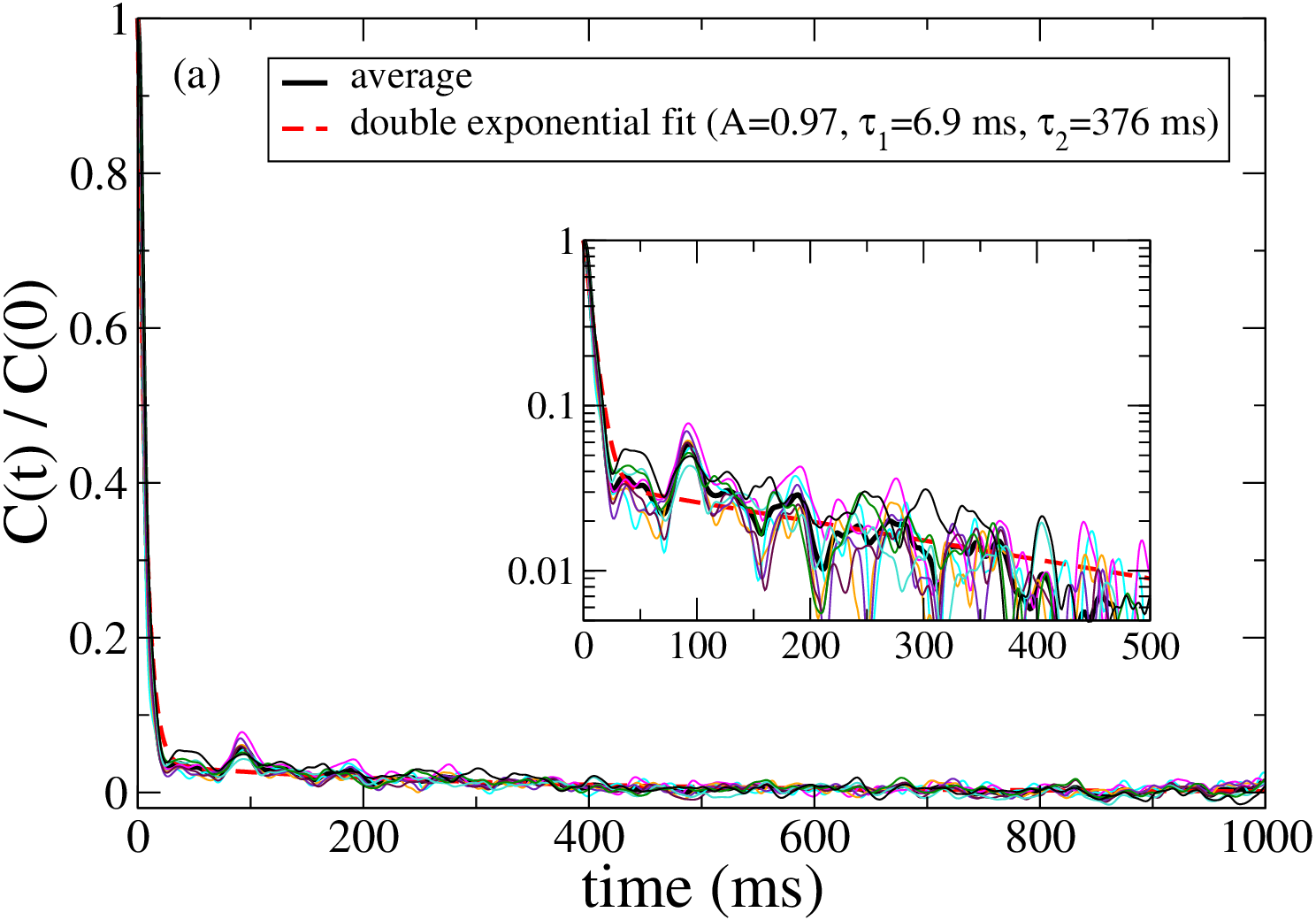}
\includegraphics[width=0.9\columnwidth,clip=true]{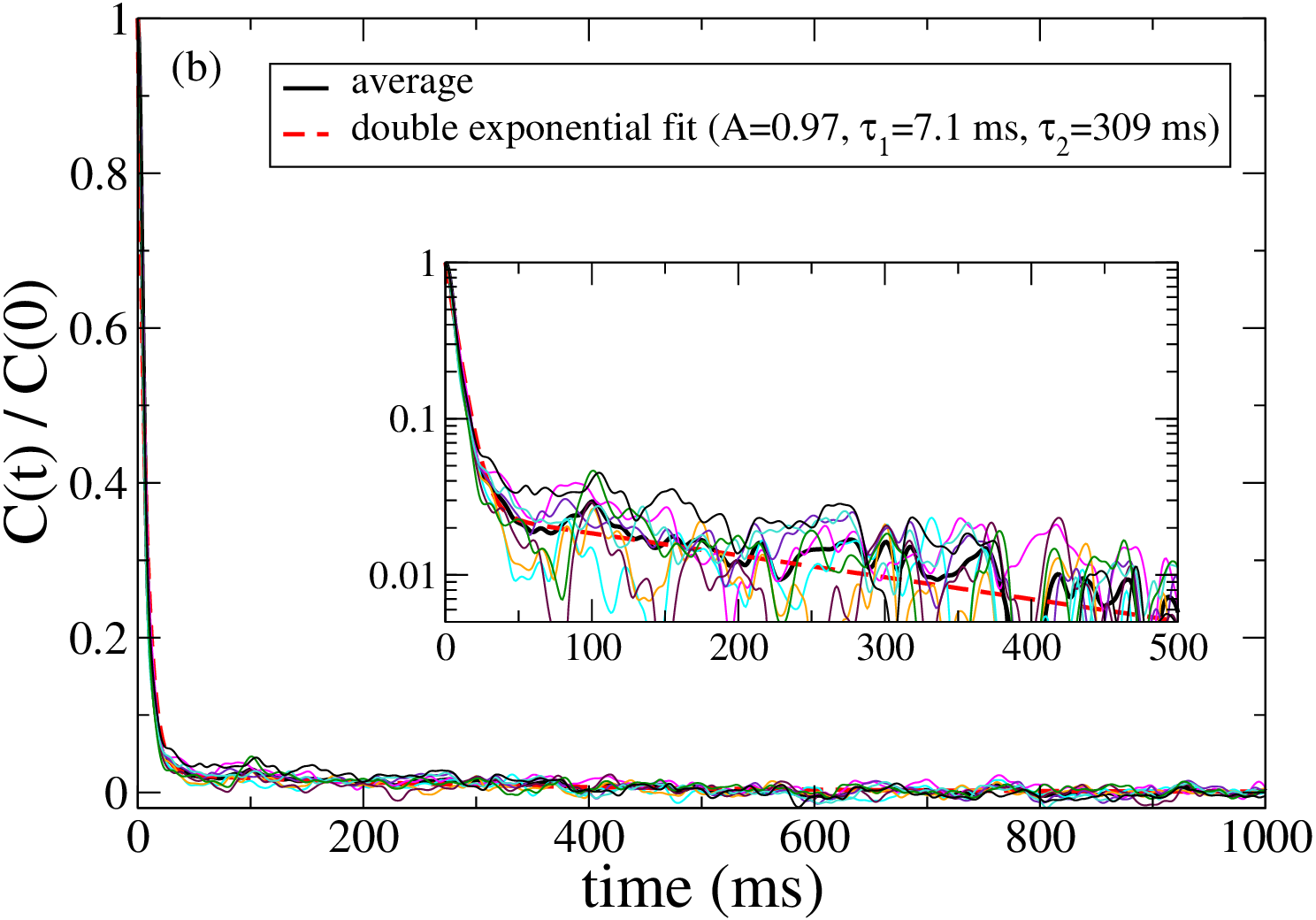}
\caption{Autocorrelation function for the spontaneous activity at eight sensors in the occipital (a) and temporal (b) areas of a single subject: Thin lines show data for each single sensor, whereas the thick black line is the average over the eight sensors.}
\label{occ_temp}
\end{figure}

\begin{figure}[!t]
\centering
\includegraphics[width=0.9\columnwidth,clip=true]{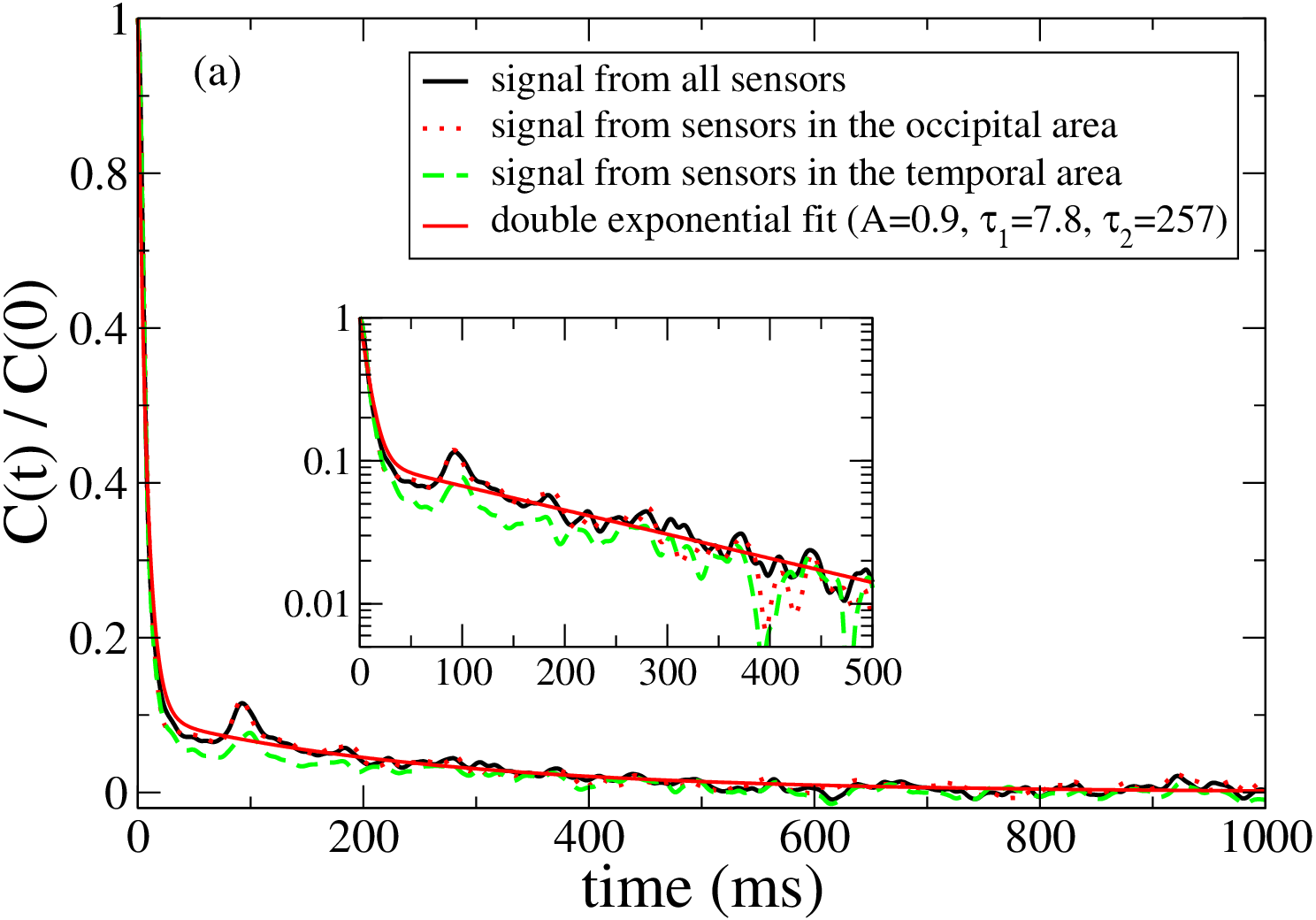}
\includegraphics[width=0.9\columnwidth,clip=true]{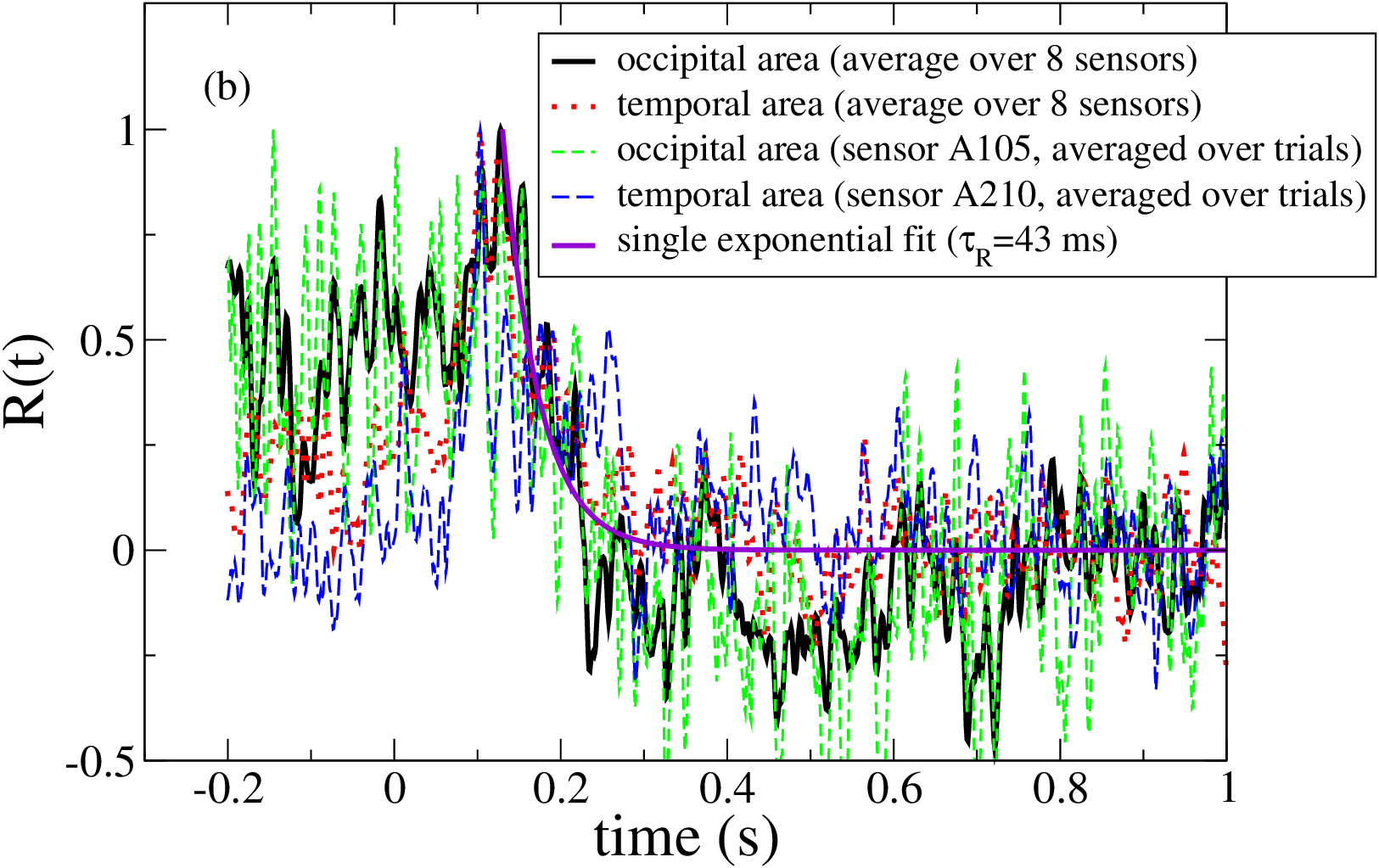}
\caption{(a) The autocorrelation function calculated for the
  global signal for each area. (b) The response function for the average signal
  from the temporal and occipital areas (thick lines, averaged over
  the number of trials), as well as for signals at two representative
  sensors in each area.}
\label{sensors}
\end{figure}

In figures~\ref{occ_temp} and~\ref{sensors}, we show the results for
the autocorrelation and the response function compared with results
for global data, for one subject (similar behavior was verified also
for the other six subjects).  In Figs.~\ref{occ_temp} thin lines show
data for each single sensor, whereas the thick black line is the
average over the eight sensors. Data show that the autocorrelation
function (in linear and log-linear scale in the inset) calculated at a
single sensor both in the occipital and temporal area and
averaged over eight sensors of the same area, confirm the double
exponential decay with characteristic times similar to the ones for
global activity.  Moreover, in Fig.~\ref{sensors} (left panel),
the autocorrelation function calculated for the global signal for each
area displays the same functional behavior with the same
characteristic times found for the global signal.

Fig.~\ref{sensors} (bottom panel) shows the response function for the
average signal from the temporal and occipital areas (thick lines,
averaged over the number of trials), as well as for signals at two
representative sensors in each area. In this case, data for each
sensor are, as expected, very noisy in all cases, however the fit of
the thick lines is again optimal with a single exponential decay where
the characteristic time (43ms for the temporal area) is in good
agreement with $\tau_R$ found for the global signal.

\section{Details on analytical computations}

The Wilson-Cowan model in the linear noise approximation consists of the linear system
\begin{equation}
  \dot{\boldsymbol{\xi}}={\cal M}\boldsymbol{\xi}+\sqrt{\alpha\Sigma_0}\boldsymbol{\eta},
  \label{system}
  \end{equation}
where $\boldsymbol{\xi}=(\xi_\Sigma,\xi_\Delta)^T$,
\begin{equation}
{\cal M}=\left( \begin{array}{cc}
    -1/\tau_1 & w_{ff} \\ 0 & -1/\tau_2 \end{array} \right),
\end{equation}
and $\boldsymbol{\eta}=(\eta_\Sigma,\eta_\Delta)^T$ are independent
delta-correlated noises, with zero average and unit variance.
The general solution of Eq.~(\ref{system}) is
\begin{equation}\label{soluz}
\boldsymbol{\xi}(t)=e^{{\cal M} t}\boldsymbol{\xi}(0)+\sqrt{\alpha \Sigma_0}\int_0^t e^{{\cal M}(t-t')}\boldsymbol{\eta}(t')dt'.
\end{equation}
The covariance matrix at stationarity $\sigma$, whose matrix elements
are the equal-time correlations $\sigma_{ij}=\langle
\xi_i(0)\xi_j(0)\rangle$ with $i,j=(\Sigma,\Delta)$, where
$\langle\cdots\rangle$ denotes an average in the stationary state,
satisfies the matrix relation~\cite{risken}
\begin{equation}
-\left( \begin{array}{cc}
  \alpha\Sigma_0 & 0 \\ 0 &   \alpha\Sigma_0 \end{array} \right)={\cal M} \sigma+\sigma {\cal M}^T,
\label{dsa}
\end{equation}
where ${\cal M}^T$ denotes the transpose matrix.
Solving Eq.~(\ref{dsa}) one obtains
\begin{equation}
  \sigma=\frac{\alpha\Sigma_0}{2}\left( \begin{array}{cc}
    \tau_1\left(1+\frac{w_{ff}^2\tau_1\tau_2^2}{\tau_1+\tau_2}\right) &
    \frac{w_{ff}\tau_1\tau_2^2}{\tau_1+\tau_2}\\ \frac{w_{ff}\tau_1\tau_2^2}{\tau_1+\tau_2} &
    \tau_2 \end{array} \right).
  \end{equation}
The inverse matrix $\sigma^{-1}$ then 
reads
\begin{eqnarray}
  \sigma^{-1}&=&\frac{2}{\alpha\Sigma_0}\frac{\tau_1+\tau_2}{2\tau_1\tau_2+\tau_2^2+\tau_1^2(1+w_{ff}^2\tau_2^2)}\nonumber \\
  &\times& \left( \begin{array}{cc}
    \frac{\tau_1+\tau_2}{\tau_1} &
    -w_{ff}\tau_2\\ -w_{ff}\tau_2 &
    \frac{\tau_1+\tau_2++w_{ff}^2\tau_1\tau_2^2}{\tau_2} \end{array} \right).
  \end{eqnarray}
The elements of the time correlation matrix ${\cal C}(t)$ are obtained from the equations~\cite{risken}
\begin{equation}
  {\cal C}_{ij}(t)=(e^{{\cal M}t}\sigma)_{ij}.
  \label{cems}
  \end{equation}
The matrix ${\cal M}$ has eigenvalues $(-1/\tau_1,-1/\tau_2)$
and eigenvectors $(1,0)^T$ and $(-w_{ff}\tau_1\tau_2/(\tau_1-\tau_2),1)^T$.
Diagonalizing ${\cal M}$, one  obtains the matrix exponential
\begin{eqnarray}
e^{{\cal M }t}&=&\left( \begin{array}{cc}
    1 &
    -\frac{w_{ff}\tau_1\tau_2}{\tau_1-\tau_2}\\ 0 &
    1 \end{array} \right)\left( \begin{array}{cc}
    e^{-t/\tau_1} &
    0\\ 0 &
    e^{-t/\tau_2} \end{array} \right) \nonumber \\
    &\times&\left( \begin{array}{cc}
    1 &
    -\frac{w_{ff}\tau_1\tau_2}{\tau_1-\tau_2}\\ 0 &
    1 \end{array} \right)^{-1}\nonumber \\
    &=&\left( \begin{array}{cc}
    e^{-t/\tau_1} &
    \frac{\tau_1\tau_2w_{ff}(e^{-t/\tau_1}-e^{-t/\tau_2})}{\tau_1-\tau_2} \\ 0 &
    e^{-t/\tau_2} \end{array} \right).
  \end{eqnarray}
  We notice that in the case of equal characteristic times $\tau_1=\tau_2$ the upper right element in the above matrix takes the value $te^{-t/\tau_1}$. 
Next, computing the matrix product in Eq.~(\ref{cems}) one obtains the explicit expressions
\begin{eqnarray}
{\cal C}_{\Sigma\Sigma}(t)&=&\frac{\alpha\Sigma_0\tau_1^2\tau_2^2}{2(\tau_2^2-\tau_1^2)}\nonumber \\
&\times&\left[(\tau_1^{-1}-\tau_1\tau_2^{-2}-\tau_1w_{ff}^2)e^{-t/\tau_1}+\tau_2
  w_{ff}^2e^{-t/\tau_2}\right] \nonumber \\
\\ {\cal C}_{\Sigma\Delta}(t)&=&\frac{\alpha\Sigma_0\tau_1\tau_2^2w_{ff}}{2(\tau_1^{2}-\tau_2^{2})}\left[2
  \tau_1 e^{-t/\tau_1}- (\tau_1+\tau_2)e^{-t/\tau_2}\right] \nonumber \\
\\ {\cal C}_{\Delta\Sigma}(t)&=&\frac{\alpha\Sigma_0\tau_1\tau_2^2w_{ff}}{2(\tau_1+\tau_2)}e^{-t/\tau_2}
\\ {\cal C}_{\Delta\Delta}(t)&=&\frac{\alpha\Sigma_0\tau_2}{2}e^{-t/\tau_2}.
\end{eqnarray}
We notice that ${\cal C}_{\Delta\Sigma}(t)$ is different than zero and describes the decay
of the initial correlation $\xi_\Delta(0)\xi_\Sigma(0)$, controlled by the characteristic time
$\tau_2$.

The matrix exponential $e^{{\cal M}t}$ coincides with the response
function matrix ${\cal R}_{ij}=\overline{\delta \xi_i(t)}/\delta
\xi_j(0)$, so that
\begin{eqnarray}
{\cal R}_{\Sigma\Sigma}(t)&=&e^{-t/\tau_1} \\
{\cal R}_{\Sigma\Delta}(t)&=&\frac{\tau_1\tau_2w_{ff}(e^{-t/\tau_1}-e^{-t/\tau_2})}{\tau_1-\tau_2} \\
{\cal R}_{\Delta\Sigma}(t)&=&0 \\
{\cal R}_{\Delta\Delta}(t)&=&e^{-t/\tau_2}.
\end{eqnarray}

\end{document}